\RequirePackage{fix-cm}
\documentclass[smallextended]{svjour3}       
\smartqed  
\usepackage{graphicx}
\usepackage{epstopdf}
\usepackage{amsmath}
\usepackage{color}

\DeclareMathOperator*{\SumInt}{%
\mathchoice%
  {\ooalign{$\displaystyle\sum$\cr\hidewidth$\displaystyle\int$\hidewidth\cr}}
  {\ooalign{\raisebox{.14\height}{\scalebox{.7}{$\textstyle\sum$}}\cr\hidewidth$\textstyle\int$\hidewidth\cr}}   
  {\ooalign{\raisebox{.2\height}{\scalebox{.6}{$\scriptstyle\sum$}}\cr$\scriptstyle\int$\cr}}
  {\ooalign{\raisebox{.2\height}{\scalebox{.6}{$\scriptstyle\sum$}}\cr$\scriptstyle\int$\cr}}  
}

\begin{document}

\title{Update on nuclear structure effects in light muonic atoms
}

\author{Oscar Javier Hernandez \and
        Nir Nevo Dinur \and
          Chen Ji \and
        Sonia Bacca \and
        Nir Barnea
}

\institute{O. Javier Hernandez \at
              TRIUMF, 4004 Wesbrook Mall, Vancouver, BC V6T 2A3 \\
              \email{javierh@triumf.ca}           
}

\date{Received: date / Accepted: date}

\maketitle

\begin{abstract}
We present calculations of the nuclear structure corrections to the Lamb shift in light muonic atoms, using state-of-the-art nuclear potentials. We outline updated results on finite nucleon size contributions.
\end{abstract}

\section{Introduction}
\label{intro}
The root-mean-square (RMS) charge radius of the proton $r_{p}$ has been extracted to unprecedented precision from recent measurements of the Lamb shift (2S-2P) transition in muonic hydrogen $\mu$H. The extracted  $r_{p}$ value is smaller by 7$\sigma$ with respect to the CODATA value, which is based on $e$H spectroscopy and $e$-$p$ scattering \cite{Phol_2010,Antognini_2013}. This unresolved discrepancy between the electron-based and the muon-based values is known as ``the proton radius puzzle", and challenges our understanding of physics based on the standard model. The CREMA collaboration at PSI has undertaken the task of extracting the nuclear RMS charge radius $R_{c}$ from measurements of the Lamb shift in several hydrogen-like muonic systems, with the aim to improve the accuracy in the determination of $R_{c}$. These new measurements may shed light on the puzzle by studying whether the discrepancy between the nuclear charge radii measured using electrons and muons persists or varies as a function of the atomic mass $A$ and charge number $Z$. For $A \geq 2$ these measurements are affected by nuclear structure corrections. Ultimately, these corrections determine the attainable precision of $R_{c}$, and it is therefore important to obtain the most accurate nuclear structure corrections with reliable uncertainties.\\ 
\begin{figure*}
\centering
  \includegraphics[scale=0.35]{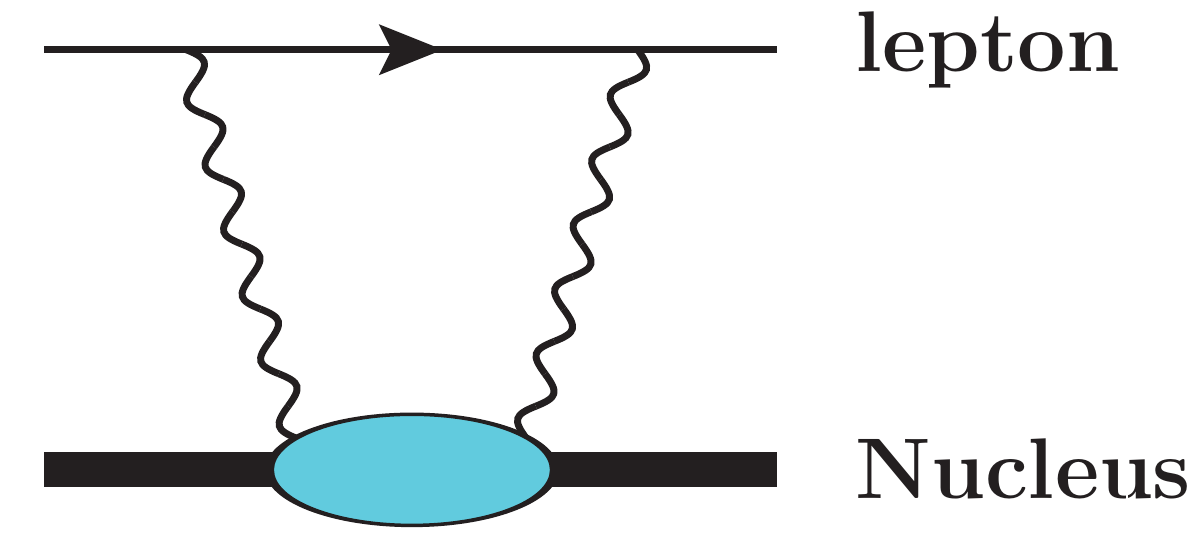}
\caption{The muon-nucleus two-photon-exchange. The bulb denotes the excitation of the nucleus in the intermediate states between the two photons.}
\label{fig:1}       
\end{figure*}

The Lamb shift is related to $R_{c}$ by
\begin{equation}
\label{eq: Lamb Shift Correction}
\Delta E = \delta_{\rm QED}+\delta_{\rm FS}(R_{c})+\delta_{\rm TPE}.
\end{equation}
The three terms, given in order of decreasing magnitude, are: the QED contributions from vacuum polarization, lepton self energy, and relativistic recoil; the leading correction due to the finite size of the nucleus, $\delta_{\rm FS}(R_{c}) = \frac{m^{3}_{r}(Z\alpha)^{4}}{12}R^{2}_{c}$ (in $\hbar=c=1$ units), from one-photon-exchange; and the two-photon exchange (TPE) contribution depicted in Fig.~\ref{fig:1}. The contributions from this diagram are divided into the elastic Zemach term and the inelastic polarization term, {\it i.e.}, $\delta_{\rm TPE} = \delta_{\rm Zem}+\delta_{\rm pol}$, which can be further separated into nuclear $(\delta^{A})$ and nucleonic $(\delta^{N})$ components, {\it i.e}, $\delta_{\rm TPE} = \delta^{A}_{\rm Zem}+\delta^{A}_{\rm pol}+\delta^{N}_{\rm Zem}+\delta^{N}_{\rm pol}$. In Refs.~\cite{Ji_2013,Hernandez_2014,Ji_2014,Nevo_Dinur_2016} we have provided values for the various terms in $\delta_{\rm TPE}$ for the hydrogen-like muonic systems with $2 \leq A \leq 4$, based on {\it ab initio} calculations with various state-of-the-art models for the nuclear Hamiltonian and newly developed methods \cite{Nevo_Dinur_2014}. This has enabled us to provide improved precision and uncertainty estimates compared to previous results. Here we provide details on recent recalculations of $\delta^{N}_{\rm Zem}$ and the corrections to $\delta^{A}_{\rm pol}$ due to the finite size of the nucleons.

\section{Nucleon size corrections to the nuclear polarizability} \label{section: NS corr to nucl pol}

The calculation of $\delta^{A}_{\rm pol}$ is portrayed in Refs.~\cite{Ji_2013,Hernandez_2014}, where the various terms contributing to $\delta^{A}_{\rm pol}$ are obtained in a multipole expansion formalism
\begin{align}
\delta^{A}_{\rm pol} &= \delta^{(0)}+\delta^{(1)}+\delta^{(2)}+\delta^{(1)}_{NS}+\delta^{(2)}_{NS}.
\end{align}
The terms $\delta^{(0)},\delta^{(1)},\delta^{(2)}$ denote the leading, sub-leading, and 2nd-order corrections, respectively, all calculated in the point-nucleon limit. The terms $\delta^{(1)}_{NS}$ and $\delta^{(2)}_{NS}$ denote the respective nucleon size (NS) corrections. The details of the calculation and formula for each of these terms are illustrated in Ref.~\cite{Ji_2013}. Some of the terms that contribute to the NS corrections have been updated in Ref.~\cite{Nevo_Dinur_2016}; below we provide more details as well as further updates.  

When the finite sizes of the nucleons are considered, the Coulomb interaction with the point-protons is substituted by an integration over the internal charge distributions of the nucleons. 
In Refs.~\cite{Ji_2013,Hernandez_2014,Ji_2014} the NS corrections were derived in momentum-space, using the low $Q^{2}$ approximation for the nucleon form factors. They were re-derived in Ref.~\cite{Nevo_Dinur_2016} in coordinate-space, as was done by Friar in Ref.~\cite{Friar_2013}. The results of this treatment are identical to those of Refs.~\cite{Ji_2013,Hernandez_2014,Ji_2014}, with the exception of a term that vanishes in the momentum-space treatment used there. This term is related to the 3rd Zemach moments of the nucleons. The contribution from the neutrons is negligible, and the contribution from the protons is estimated from the corresponding term in muonic hydrogen $\delta_{Zem}(\mu {\rm H})$ according to
\begin{equation}
\label{eq: Zemach}
\delta^{N}_{\rm Zem}(\mu A) = \left(\frac{Zm_{r}(\mu A)}{m_{r}(\mu H)} \right)^{4}  \times \delta_{\rm Zem}(\mu H).
\end{equation}
The scaling factor in the above equation can be extracted from Ref.~\cite{Friar_2013}. However, it differs from the one used, e.g., in Ref.~\cite{Krauth_2016} for both the elastic and inelastic nucleon TPE, and calls for further investigation.

Next we detail the 2nd order NS corrections $\delta^{(2)}_{\rm NS}$. The neutron-neutron contribution is negligible, therefore only proton-proton $\delta^{(2)}_{pp}$ and neutron-proton $\delta^{(2)}_{np}$ terms contribute. The former is expressed, as in Ref.~\cite{Ji_2013}, by 
\begin{equation}
\delta^{(2)}_{pp} = -\frac{8\pi}{27}m^{5}_{r}(Z\alpha)^{5}r^{2}_{p}\int\limits_{\omega_{th}}^{\infty}d\omega \sqrt{\frac{\omega}{2m_{r}}} S_{D_{1}}(\omega), \label{eq: deltaPP proton-proton}
\end{equation}
 using the dipole response
\begin{align}
S_{D_{1}}(\omega) = \SumInt\limits_{f\neq 0}|\langle f |\hat{D}_{1}| 0 \rangle|^{2}\delta(E_{f}-E_{i}-\omega), \label{eq:Dipole Response}
\end{align}
where $\hat D_1 = \frac{1}{Z} \sum_{a}^{A} r_a Y_1(\hat r_a) \frac{1+\tau_a^3}{2}$, with $Y_1$ the rank-1 spherical harmonic. The latter, however, takes the form
\begin{align}
\label{eq:deltaNS neutron-proton}
\delta^{(2)}_{np} &= {-\frac{8\pi}{27}} m_r^5 (Z \alpha )^5 \frac{N}{Z}r^2_n
\int^\infty_{\omega_{\rm th}}
d\omega \sqrt{\frac{\omega}{2m_r}} S_{D_p D_n}(\omega)\notag \\
&={\ \ \frac{8\pi}{27}} m_r^5 (Z \alpha )^5 r^2_n
\int^\infty_{\omega_{\rm th}}
d\omega \sqrt{\frac{\omega}{2m_r}} S_{D_1}(\omega),
\end{align}
where
 \begin{equation}
\label{eq:Sd1 and Sdpdn definition}
S_{D_p D_n}(\omega) =  \SumInt\limits_{f\neq 0}\langle 0 |\hat{D}^{\dagger}_{n}| f \rangle \langle f |\hat{D}_{p}| 0 \rangle\delta(E_{f}-E_{i}-\omega)
\end{equation}
contains the protonic dipole operator $\hat{D}_{p}=\hat{D}_{1}$ and the neutronic dipole operator $\hat D_n = \frac{1}{N} \sum_{a}^{A} r_a Y_1(\hat r_a) \frac{1-\tau_a^3}{2}$. The isoscalar part of both operators is proportional to the nuclear center of mass motion, and vanishes between the nuclear ground and excited states. Thus, we obtain $S_{D_p D_n}(\omega) = -(Z/N)S_{D_1}(\omega)$, which was used in the last step of Eq.~\eqref{eq:deltaNS neutron-proton}. The expression in Eq.~\eqref{eq:deltaNS neutron-proton} supersedes the corresponding ones in \cite{Ji_2013,Nevo_Dinur_2016}. The results in Eqs.~\eqref{eq: deltaPP proton-proton} and \eqref{eq:deltaNS neutron-proton} coincide with the corresponding expression for the special case of deuterium given in Ref.~\cite{Pachucki_2015}, and were used in a recent summary of the theory related to the $\mu$D experiment \cite{Krauth_2016}.

\section{Results}
Integrating the new $\delta^{N}_{\rm Zem}$ and $\delta^{(2)}_{\rm NS}$ discussed above into the results of Refs.~\cite{Ji_2013,Hernandez_2014,Nevo_Dinur_2016} we obtain updated values for $\mu$D, $\mu^{4}{\rm He}^{+}$, $\mu^{3}{\rm He}^{+}$, and $\mu^{3}{\rm H}^{+}$. In Table~\ref{tab:1} below we present updated results for the individual terms in $\delta_{\rm TPE}$ for all $2 \leq A \leq 4$ hydrogen-like muonic systems, where the results for $A=2,4$ are taken from Refs.~\cite{Ji_2013,Hernandez_2014,Ji_2015}, and for $A=3$ the results given here are taken from Ref.~\cite{Nevo_Dinur_2016} and supersede it due to Eq.~\eqref{eq:deltaNS neutron-proton} and a correct implementation of Eq.~\eqref{eq: Zemach}. More details will be given in Ref.~\cite{Ji_et_al}.
For completeness we include $\delta^{N}_{\rm pol}$ in Table \ref{tab:1}, where for $\mu$D its value and uncertainty were extracted from data in Ref.~\cite{Carlson_2014}, and for other systems they were estimated based on their values in $\mu$H, as explained in Ref.~\cite{Nevo_Dinur_2016}.
\begin{table}[htb]
\centering
\caption{Contributions to $\delta_{\rm TPE}$ of the Lamb shift in light muonic atoms, in meV. The uncertainty associated with each value is given in brackets and includes the numerical, nuclear model, and atomic physics uncertainties. Due to cancellation of the Zemach term, the uncertainty in $\delta_{\rm TPE}$ can differ from the quadrature sum of the corresponding terms in the table. A detailed analysis of the uncertainties is given in Ref.~\cite{Nevo_Dinur_2016}.}
\label{tab:1}       
\begin{tabular}{l|llll|l}
\hline\noalign{\smallskip}
 & $\delta^{A}_{\rm Zem}$ & $\delta^{A}_{\rm pol}$ & $\delta^{N}_{\rm Zem}$ & $\delta^{N}_{\rm pol}$ & $\delta_{\rm TPE}$  \\
\noalign{\smallskip}\hline\noalign{\smallskip}
$\mu$D & -0.424(3) &  -1.245(19) & -0.030(2) & -0.028(2) & -1.727(20) \\
$\mu ^{3}\text{H}$ & -0.227(6) & -0.473(17) & -0.033(2) & -0.034(16) &  -0.767(25)  \\
$\mu ^{3} \text{He}^{+}$ & -10.49(24) &  -4.17(17)& -0.52(3)  & -0.28(12) & -15.46(39) \\
$\mu  ^{4}\text{He}^{+}$ & -6.29(28) & -2.36(14) & -0.54(3) & -0.38(22) &   -9.58(38) \\
\noalign{\smallskip}\hline
\end{tabular}
\end{table}

The results in Table \ref{tab:1} provide rigorous estimates of the nuclear structure corrections to the Lamb shift in the lightest hydrogen-like muonic systems, and a careful treatment of their associated uncertainties. Together with the highly precise determinations of the other terms in Eq.~\eqref{eq: Lamb Shift Correction}, these results will allow the nuclear charge radii of these isotopes to be determined with significantly improved precision. This development should be insightful not only towards the resolution of the proton radius puzzle but also to probe possible effects due to the neutrons.

\end{document}